# Informing Additive Manufacturing technology adoption: total cost and the impact of capacity utilisation

Martin Baumers*, Luca Beltrametti†, Angelo Gasparre†, and Richard Hague*

*\* Centre for Additive Manufacturing, University of Nottingham, Nottingham, NG7 2RD, United Kingdom*

*† Department of Economics and Business, University of Genoa, Via F. Vivaldi 5, 16126 Genoa, Italy*

Correspondence to: Martin Baumers, Faculty of Engineering, University of Nottingham, Nottingham, NG7 2RD, United Kingdom. Email: martin.baumers@nottingham.ac.uk, Phone: +44 (0)115 951 3877.

# Informing Additive Manufacturing technology adoption: total cost and the impact of capacity utilisation

Informing Additive Manufacturing (AM) technology adoption decisions, this paper investigates the relationship between build volume capacity utilisation and efficient technology operation in an inter-process comparison of the costs of manufacturing a complex component used in the packaging industry. Confronting the reported costs of a conventional machining and welding pathway with an estimator of the costs incurred through an AM route utilising Direct Metal Laser Sintering (DMLS), we weave together four aspects: optimised capacity utilisation, ancillary process steps, the effect of build failure, and design adaptation. Recognising that AM users can fill unused machine capacity with other, potentially unrelated, geometries, we posit a characteristic of "fungible" build capacity. This aspect is integrated in the cost estimation framework through computational build volume packing, drawing on a basket of sample geometries. We show that the unit cost in mixed builds at full capacity is lower than in builds limited to a single type of geometry; in our study this results in a mean unit cost overstatement of 157%. The estimated manufacturing costs savings from AM adoption range from 36% to 46%. Additionally, we indicate that operating cost savings resulting from design adaptation are likely to far outweigh the manufacturing cost advantage.

**Introduction**

Additive Manufacturing (AM), also known as 3D Printing, has seized the imagination of many manufacturing professionals and technology experts. The technology is regarded as a pathway to digitise production, manufacture on demand and rethink product design. While the technological characteristics of AM have been subject to investigation from different perspectives, a more detailed and realistic grasp of the business case towards AM adoption is needed.

Besides facing a number of technological limitations and organisational challenges (Khorram Niaki and Nonino, 2017), AM processes are associated with two advantages over conventional manufacturing techniques (Tuck et al., 2008; Beltrametti and Gasparre, 2016; Gardan, 2016). Firstly, AM allows the user to ignore numerous tooling-related constraints that are placed on product geometries through conventional manufacturing processes (cf. Boothroyd and Dewhust, 1994). The "free form" characteristic of AM, in fact, allows the manufacturing of goods with an intrinsically better technical and functional profile. Secondly, AM enables the efficient manufacture at very low volumes, potentially down to a single unit, thereby enabling the manufacture of customised or highly differentiated products (Eyers et al., 2008; Berman, 2012) and resulting in novel operational practices (Holmström et al., 2017). We see these aspects as responsible for the euphoria in those who evoke a new AM-driven paradigm of sweeping mass customization and spare parts production on demand, thus reducing warehousing and capital costs (Achillas et al., 2017). Seminal work in cost modelling for AM was carried out by Alexander et al. (1998), falling under the broad category of activity-based costing. Such models assign direct and indirect costs arising from resource consumption to identifiable activities in the business context (cf. Niazi et al., 2006). In the case of AM cost models, indirect costs are allocated through the duration of various processes occurring in the AM work flow.

The cost model by Alexander et al. assembled time estimates for multiple process steps, including the duration of the AM build process itself. This approach can be reconfigured for different AM operating systems; it can also be adapted to different process settings by changing the tasks surrounding the core build AM build process, for example incorporating pre- and post-processing costs, or be extended to include the cost impact of different supply chain configurations (Li et al., 2017). As noted by Rickenbacher et al. (2013), the approach taken by Alexander et al. is limited to the assessment of the cost of individual products in isolation. It will therefore not be valid for builds containing multiple parts, which is the normal pattern of operation for most AM systems (Ruffo and Hague, 2007). Hopkinson and Dickens (2003) investigated the utilisation of AM for larger quantities and at high levels of capacity utilisation, effectively indicating that the quantity-unit cost relationship, central to conventional manufacturing, may be absent in AM.

The cost model by Ruffo and Hague (2007) showed that the degree of capacity utilisation determines the average unit cost for major AM technology variants; this result has been replicated for process energy consumption (Baumers et al., 2011). Moreover, it has been shown by Ruffo and Hague that composing builds of dissimilar parts equally affects unit costs, suggesting that technology users face a non-trivial problem of assembling cost minimising build configurations in practise. This aspect has previously been incorporated in an AM cost model by Baumers et al. (2013), integrating computational build volume packing within a cost estimation framework.

Moreover, it is understood in engineering that the selection of design, process, and material are interdependent (cf. Reuter, 2007). This implies that unilateral change in any of these areas is likely to produce an invalid configuration – providing grounds for criticism of approaches choosing a technology through the identification of break-even quantities (e.g. Hopkinson and Dickens, 2003; Ruffo and Hague, 2007; Achillas et al, 2017). Atzeni et al. (2010) and Atzeni and Salmi (2012) have addressed this issue by comparing the cost of a tooled manufacturing process against an AM pathway for a redesigned component capable of performing an equivalent function.

A further aspect of practical relevance that has so far been largely omitted in AM cost modelling is the cost impact of the risk of build failure, which could be classified as an ill-structured cost (Son, 1991). In order to accommodate such facets, quality control systems have been included in AM cost investigations (Schmid and Levy, 2014, Berumen et al., 2010). Baumers and Holweg (2016) addressed this aspect directly by modelling the expected cost impact of different failure modes.

To inform the justification of AM technology adoption, we develop a novel methodology for cost estimation in AM by pulling together four salient characteristics identified in the literature on AM cost estimation within a single activity-based costing framework. Centring on an ability to freely fill available machine capacity, to which we will return later in the paper, we identify the problem of configuring the available build space and show how this problem can be addressed for real world settings by accommodating the following aspects:

(1) In reality AM takes place within a series of process steps. Thus, an inter-process comparison must reflect the process maps observed in industry and acknowledge that, as shown by Mellor et al. (2014), AM processes do not take place in isolation.
(2) By mimicking the commercial practise of mixing build of multiple, potentially unrelated geometries, this paper takes a realistic view on how AM is used (Ruffo and Hague, 2007). This is achieved by utilising a computational build volume packing approach and a set of reference parts to synthetically build fill up empty capacity (cf. Araujo et al., 2015).
(3) Recent research has shown that the risk of build failure places a great burden on the real cost of using AM, as shown by Baumers and Holweg (2016) for polymeric Laser Sintering. To address this aspect, this paper integrates a simple model of AM build failure within the cost estimator.
(4) By accepting that the valid designs and material grades are tied to manufacturing process selection in reality (Reuter, 2007), the paper avoids the problem of comparing technically inappropriate designs. This is achieved by treating product designs as given and specific to processes. Additionally, we provide a perspective on use-phase consequences of design modification. It should be noted that we concentrate on manufacturing cost and hence ignore the cost of re-design.

To maintain a firm empirical footing, which is prevalent in theory-building research of AM management (Khorram Niaki and Nonino, 2017), our framework is applied to a case study from the food industry. We assess a stainless steel hot air blower, which is a key component used in final folder module for a food packaging machine. As shown in Figure 1, this component is considered in the paper in two different versions. Firstly, we report the cost of a conventional component, shown in Figure 1a, manufactured in a combination of Computer Numerically-Controlled (CNC) machining, turning and Tungsten Inert Gas (TIG) welding. This is compared against our estimates of the cost of an AM pathway for a redesigned component, manufactured on the AM technology variant Direct Metal Laser Sintering (DMLS), as shown in Figure 1b.

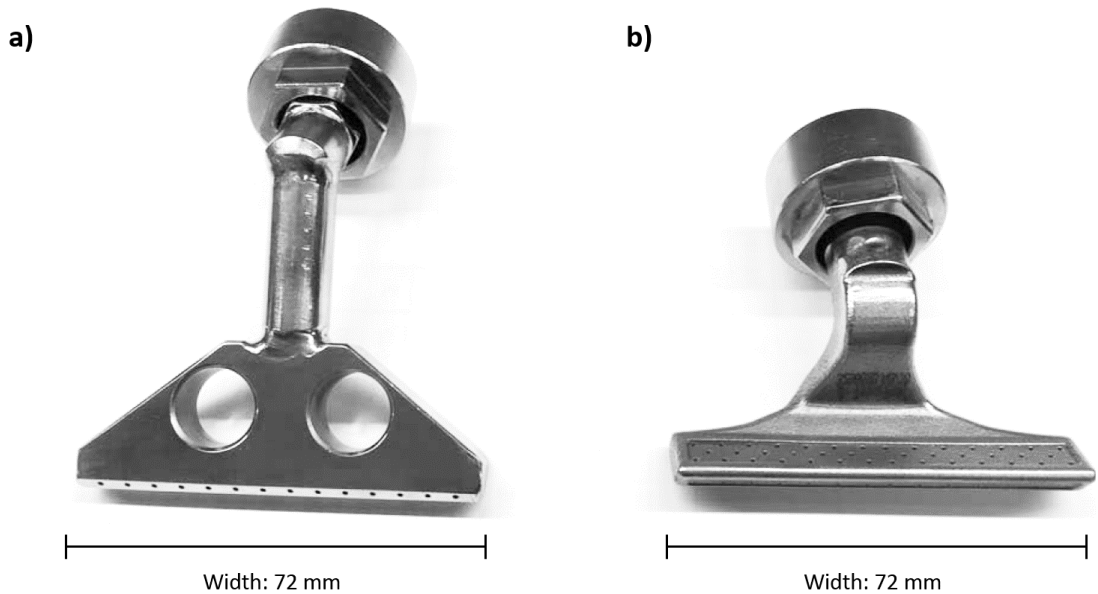

Width: 72 mm    Width: 72 mm

**Figure 1: Blower component manufactured via a conventional pathway (a) and via AM (b)**

In the AM pathway, the blower is manufactured using an EOSINT M270 system, which builds up metal components through the sequential deposition of thin horizontal layers (EOS GmbH, 2016). Each layer is processed by selectively melting the surface of a metal powder bed using a 200W fibre laser and then depositing a fresh increment of powder. This cycle is repeated until the build is complete. To allow the dissipation of energy into the machine frame and to stop the deposited material from deforming, DMLS requires all parts to be connected to a removable substrate through sacrificial anchor structures. The finished parts are removed from this build plate following the AM process through a wire erosion procedure.

The following section presents the methodology employed for the construction of the enhanced activity based-costing methodology and summarises the empirical data collected for this research. The subsequent section executes the cost model and presents the cost estimation results with a focus on the effects of capacity utilisation. The discussion section evaluates and contextualises the reached results in the literature. Conclusions are drawn in a final section.

**Methodology**

*Process maps for conventional manufacturing and AM*

Both versions of the investigated blower component shown in Figure 1 can be used in the final folder module supplied to the manufacturer of the packaging machine, Company A, by the manufacturer of the folding module, Company B. However, as Company B does not operate AM technology in-house, the AM version (Figure 1b) is manufactured by a specialist provider of AM services, Company C.

In the conventional pathway, Company B fabricates three stainless steel (AISI grade 304L) components using a turning process and 4-axis CNC machining. These components are subsequently joined via TIG welding. We have obtained cost data from Company B and do not construct a cost model for these processes. As the processes following the welding step are shared with the AM pathway, we draw the boundary of our analysis at this point.

In the AM route, the manufacturing of the blower through DMLS is subcontracted to Company C. Apart from procuring the stainless steel 17-4PH raw material, Company C carries out a number of post-processing and inspection steps before the blower is transferred to Company B. This analysis investigates the cost effect of potential build failure in the AM pathway by placing a build failure node in this process map, thereby determining which process steps will have to be repeated if the build operation fails.

Following both manufacturing pathways, and outside of the boundary of our investigation, the part is electropolished (to a surface roughness of $R_a$=3.2) after which a connecting brass sleeve and nut are attached by TIG welding. The functional module containing the final folder is then assembled and transferred to Company A for final testing and shipping to the end user. Figure 2 graphically summarises the process chains for both manufacturing pathways, bringing together our cost model of AM with the costs for the conventional process, as reported by Company B.

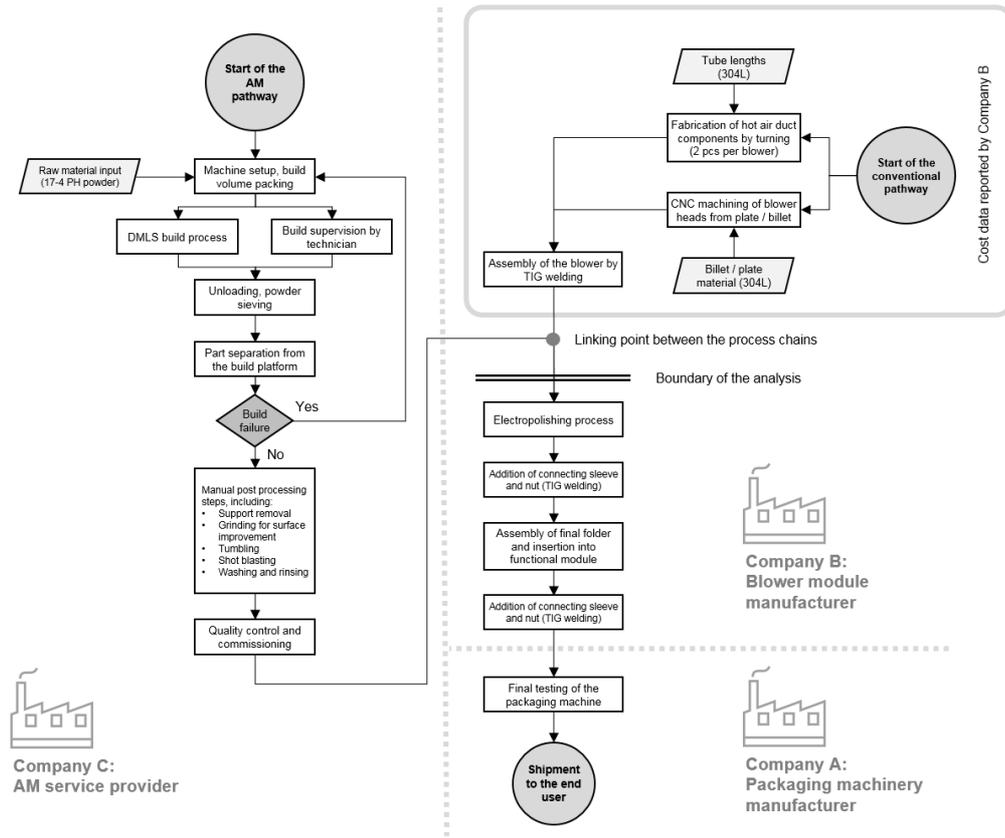

**Figure 2: Process map showing the conventional and AM pathways**

*Cost data collected for conventional manufacturing*

The total cost estimates used in this research for the conventional pathway were provided by Company B and are sensitive to batch size. For a small batch of 20 blowers a total unit cost of €105.00 is estimated. For a large batch of 60 blowers as total unit cost of €90.97 is estimated. Details of the cost elements are provided in Table 1; due to commercial sensitivity, the cost components are expressed in percentage terms.

**Table 1: Manufacturing cost estimates, conventional pathway**

| Cost item | Material / process | Cost per unit |
|---|---|---|
| Raw material costs | | |
| Tubes (2 pieces required, small quantity, 20 sets) | Stainless steel 304L | 0.95% |
| Tubes (2 pieces required, large quantity, 60 sets) | Stainless steel 304L | 1.10% |
| Billet/plate material (small quantity, 20 sets) | Stainless steel 304L | 3.81% |
| Billet/plate material large quantity, 60 sets) | Stainless steel 304L | 4.40% |
| Process costs | | |

| Fabrication of duct component (small quantity, 20 sets) | Turning | 4.76% |
|---|---|---|
| Fabrication of duct component (large quantity, 60 sets) | Turning | 4.40% |
| Machining of blower head (small quantity, 20 units) | 4-axis CNC machining | 80.95% |
| Machining of blower head (large quantity, 60 units) | 4-axis CNC machining | 82.41% |
| Welding assembly (small quantity, 20 sets) | TIG welding | 9.52% |
| Welding assembly (large quantity, 60 sets) | TIG welding | 7.69% |

The negative relationship between unit cost and quantity evident from the data provided by Company B for the machining pathway is supported by the literature on the costs of flexible manufacturing systems (Norman and Thisse, 1999; Weller et al., 2015). Associated with this, decreasing average unit costs are also associated with designated manufacturing processes employing fixed tooling (Hopkinson and Dickens, 2003; Atzeni et al., 2010; Atzeni and Salmi, 2012).

*A cost model for the AM route*

In this investigation, direct costs, incurred for raw materials (including sacrificial support structures) and energy costs are combined with indirect costs, allocated through build time $T_{Build}$ in the form of an indirect cost rate $\dot{C}_{Indirect}$, including overheads, consumables, maintenance, and machine costs. Machine preparation, build setup and unloading enters the model as a fixed labour cost increment $C_{Setup\ Labour}$. Beyond the AM process, the model considers the total post processing duration $T_{Process}$, which includes support removal, surface improvement, washing and inspection. Of course, these durations are specific to the investigated application of DMLS. The data used in the AM cost model are summarised in Table 2.

**Table 2: Cost model elements, DMLS pathway**

| Cost model element | Value |
|---|---|
| Production overhead rate, incl. rent | € 5.11 / h |
| Administration overhead rate | € 0.35 / h |
| Machine utilisation | 57.04 % |
| Annual operating hours | 5000 h |
| Machine purchase | € 411048.68 |
| Maintenance costs | € 24854.11 / year |
| Machine consumables | € 2867.78 / year |
| Wire erosion costs | € 8165.00 / year |
| Total machine cost rate | € 17.66 / h |

| Indirect cost rate, $\dot{C}_{Indirect}$ | € 23.12 | / h |
|---|---|---|
| Price for 17-4 PH material | € 88.90 | / kg |
| Price of 17-4 PH material, volume based, using an as deposited density of 7.78 g/cm$^3$, $P_{Material}$ | € 0.6916 | / cm³ |
| Energy price | € 0.02 | / MJ |
| Process energy consumption rate | 9.18 | MJ / h |
| Process energy consumption cost rate, $\dot{C}_{Energy}$ | € 0.18 | / h |
| Fixed machine setup and unloading time, per build | 180 | minutes |
| Direct Labour cost increment for machine setup and supervision, $C_{Setup\ Labour}$ | € 72.04 | |
| Total secondary processing time per part, $T_{Process}$ | 37 | minutes |
| Total production labour cost rate, $\dot{C}_{Labour}$ | € 22.75 | / h |

It should be noted that we make the simplifying assumption that no costs are incurred for equipment apart from the AM system and a wire erosion machine separating the parts from their build platform, with the costs of both machines entering through the indirect cost rate ($\dot{C}_{Indirect}$). Thus, all costs apart from indirect costs, material costs and energy costs are assumed to be labour costs, incurred at a labour cost rate $\dot{C}_{Labour}$ of €22.75 per h (Baumers and Holweg, 2016) for a total post processing duration of $T_{Process}$. Thus, the basic cost model can be specified as:

$$C_{Build} = P_{Material}V_{Build} + C_{Setup\ Labour} + (\dot{C}_{Indirect} + \dot{C}_{Energy})T_{Build} \qquad (1)$$

An additional consideration shaping the cost model developed in this paper is that AM builds are likely to be composed of multiple geometries. In reality, it is possible for AM technology users to fill the available build space with dissimilar, potentially entirely unrelated orders, as reported by Company C. Therefore, the ability to freely assign available build space to other products, which may have different sizes and shapes, needs to be reflected in the cost estimation framework. This leads to a capacity utilisation problem in which excess capacity can be filled with additional components, ensuring an optimised degree of machine utilisation, and an efficient unit cost level.

To address this problem in the context if our cost estimation methodology, it was necessary to automatically draw reference parts from a repository, configure the build volume and execute the build time model. As this functionality was not available in commercial AM workstream optimisation software such as Materialise Streamics (Materialise NV, 2017), we make use of a self-developed computational build volume packing and build time estimation tool (implemented in C++), designed to yield user-defined manufacturing configurations exploiting the available build volume capacity and to estimate $T_{Build}$ in a voxel-based implementation. The tool operates by filling the available machine capacity with blower components and additional reference parts algorithmically drawn form a basket of reference geometries. This methodology has been described and previously validated by Baumers et al. (2013).

Figure 3 shows the investigated blower component with the sacrificial anchoring structures required for DMLS in isolation (Figure 3a), an in a mixed, algorithmically packed build composed of four blower units and a number of additional reference parts (Figure 3b). The four types of reference parts algorithmically inserted are reflective of products manufactured on the investigated DMLS system (shown in Figure 3c).

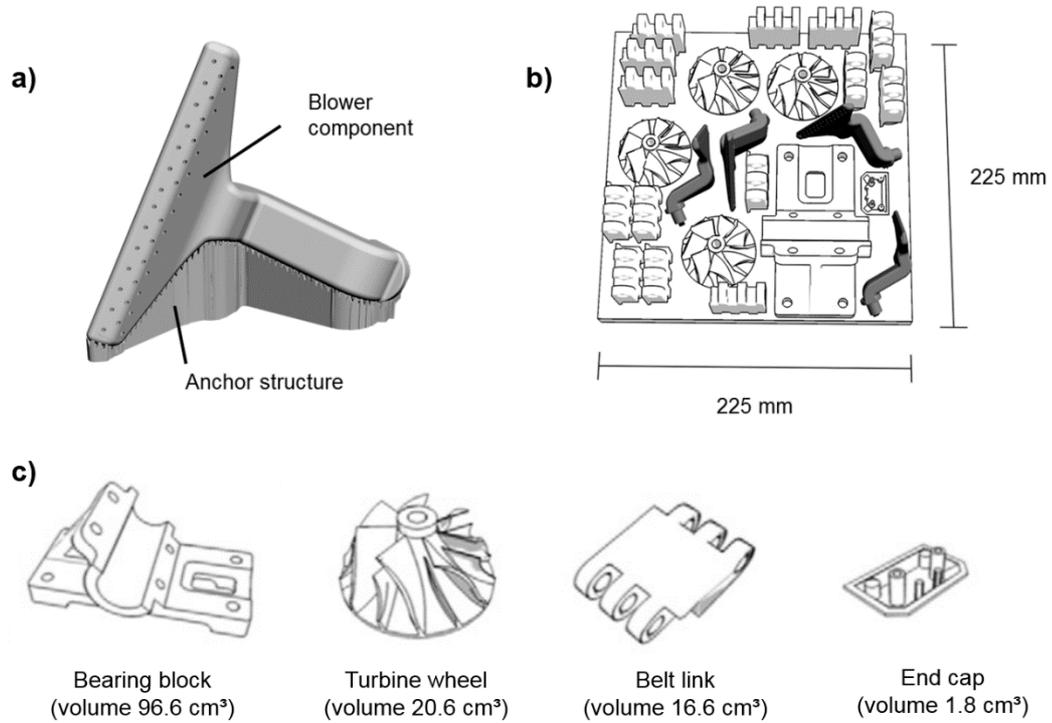

**Figure 3: Blower in build orientation with supports (a), in the build context (b), reference parts (c)**

Once a build has been composed and build time has been estimated computationally, it is possible to apportion the unit cost of the each blower through the volume fraction $v$. Each blower has a volume $V_{Blower}$ = 8.403 cm³ (including sacrificial anchor structures) and the build has a total deposited volume $V_{Build}$, so that:

$$v = \frac{V_{Blower}}{V_{Build}} \quad (2)$$

As described in Table 2, the durations for post processing $T_{Process}$ have been provided by Company C on a per-unit level. Hence, the unit cost model $C_{Unit}$ can be obtained by breaking down $C_{Build}$ as follows:

$$C_{Unit} = vC_{Build} + \dot{C}_{Labour}T_{Process} \quad (3)$$

*Modelling the expected cost of build failure*

Following the definition of process maps and the localisation of a point at which build failure takes its effect in AM, as shown in Figure 2, a specification allowing the inclusion of al build failure parameter can be formulated. Initially, we decided to keep this specification as simple as possible by limiting it to a single build failure type. The interpretation is that any excessive deviation from intended part geometry or unrecoverable disturbance of the build process is classed simply as outright build failure, leading to the write-off of all parts contained in the build.

The approach is guided by the assumption that failure events take place with a given probability, occurring on a per-layer basis, mirroring the layer-by-layer material deposition process in AM. Therefore, this research assumes that there is an independent and constant probability of build failure during each layer deposition operation $p_{Constant}$. Of course, this approach makes a simplification by implying the absence of a relationship between geometry and build failure.

Reliable empirical information on the probability of build failure in AM is rare. For an EOS P100 polymeric Laser Sintering system, which is related to the investigated DMLS system, Baumers and Holweg (2016) report a mean number of depositable layers before build failure of 4040.75. This estimate is used to approximate the constant probability of build failure per layer $p_{Constant}$ at 0.025%. By employing a discrete probability tree model with $p_{Constant}$ and the deposition of $n$ layers, the overall probability of successfully completing the build can consequently be modelled as shown in Figure 4.

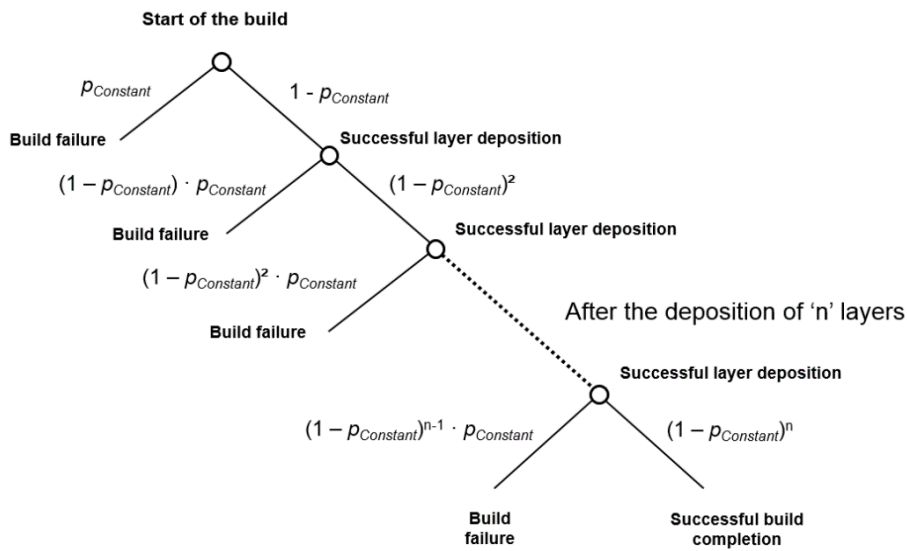

**Figure 4: Probability tree for successful build outcome based on $p_{Constant}$**

We attach the simple build failure model to the cost model by assuming that an expected cost term of manufacturing cost can be formed through the multiplication of the inverse of the probability of successful build completion $(1 - p_{Constant})^n$ with the elements of the cost model that precede the build failure node. Hence, the total unit cost model $C_{Total}$, reflecting the expected cost impact of build failure, can be expressed as:

$$C_{Total} = vC_{Build}(1 - p_{Constant})^{-n} + \dot{C}_{Labour}T_{Postprocess} \qquad (4)$$

*Capturing use phase benefits*

To avoid the dangers of constructing inter-process comparisons without considering knock-on effects on manufacturability and part design, Atzeni et al. (2010) and Atzeni and Salmi (2012) incorporated part redesign into their analysis of process economics of AM. Following in these footsteps, this research takes the position that progress resulting from the adoption of new manufacturing technologies should also manifest itself in improved product properties. Therefore, this paper performs an initial analysis of the use-phase benefits resulting from such design alteration.

The benefit of the redesign of the blower for DMLS, which has been executed by Company A, manifests itself during the product's use-phase primarily in a lower fractional process energy consumption associated with: (1) conformal, internal, channels for hot air, optimized from the point of view of fluid dynamics, and (2) a better shape of the blower that allows a more exact positioning of the devices, benefitting end product attributes (fold quality). Company A has provided two use phase scenarios for the packaging machine and estimated the energy savings resulting from the improved design. This allows the estimation of a benefit arising to the end user over the blower's useful life in monetary terms. Table 3 summarises these attributes for both the conventional part (as shown in Figure 1a) and the redesigned AM component (Figure 1b). Please note that the original component and the redesign for AM exhibit the same planned usephase duration (7.411 years) and that this is independent of operating parameters, as specified by Company A.

**Table 3: Use phase model, conventional route versus AM**

| Use phase attribute | Conventional design | Design for AM |
|---|---|---|
| Process speed (high) | 40,000 / h ||
| Process speed (low) | 8,000 / h ||
| Component lifetime | 30,000 h ||
| Component depreciation period, k | 7.411 years ||
| Annual operating hours | 4048 h/year ||
| Number of units processed annually (high) | 161,920,000 / year ||
| Number of units processed annually (low) | 32,384,000 / year ||
| Lifetime number of units processed (high) | 1,200,000,000 ||
| Lifetime number of units processed (low) | 240,000,000 ||
| Blower subsystem fractional power consumption | 3,000 W | 2,490 W |
| Annual energy consumption of blower subsystem | 43,718.40 MJ | 36,286.27 MJ |
| Annual energy cost attributable to blower | 1,692.67 € / y | 1,404.91 € / y |
| Annual saving from energy consumption reduction though the adoption of AM, $S_{Energy}$ | 287.76 €/year ||

To arrive at a present value of the projected use-phase energy saving arising from the adoption of the AM route, the discounted saving $DS_{Energy}$ resulting from energy consumption cost reduction $S_{Energy}$ accruing over the depreciation period $k$ can be modelled as a continuous annuity with discount rate r:

$$DS_{Energy} = \int_0^k S_{Energy}(1-r)^{k-t}\,dt \qquad (5)$$

which can be rearranged and evaluated to yield:

$$DS_{Energy} = S_{Energy}\frac{(1-r)^k - 1}{\ln(1-r)} \qquad (6)$$

**Results**

A useful first step in analysing the developed cost model is to determine the cost composition of the build configuration shown in Figure 3b, simulating the actual manufacturing configuration within Company C, containing four units of the blower and a number of additional reference components. Using the specification without risk of build failure, $C_{Unit}$, the shares of direct material costs and indirect costs can be identified. By subtracting $C_{Unit}$ from $C_{Total}$, the expected cost impact of build failure per unit can be stated. The cost associated with process energy consumption is obtained by multiplying the identified energy cost rate $\dot{C}_{Energy}$ with the computational estimate for build time $T_{Build}$. As labour costs enter the model both in terms of post processing and machine operation, a distinction can be made between labour costs $\dot{C}_{Labour}$ incurred for manual post processing and a fixed labour cost increment $C_{Setup\ Labour}$ for machine setup and build removal.

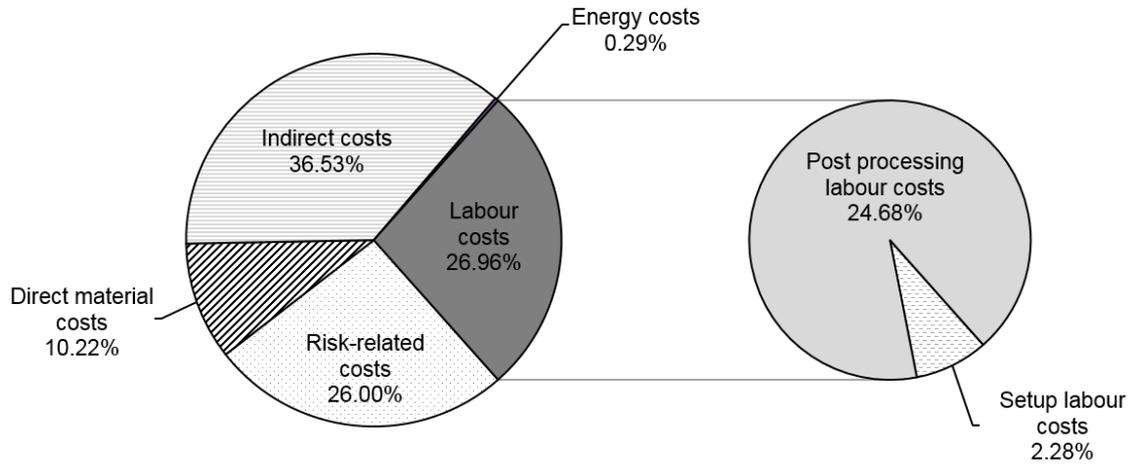

**Figure 5: Breakdown of unit cost in the real AM build configuration (as shown in Figure 3b)**

As can be seen from Figure 5, the largest item (37%) is indirect cost, relating to the DMLS machine, wire erosion, and overheads, which corresponds to the levels reported in the literature (Baumers et al., 2016; Piili et al., 2015, Atzeni and Salmi, 2012, Rickenbacher et al., 2013). The second largest cost is the risk of build failure, at 26%, which emphasises that process instability can severely affect the overall value proposition of AM. This supports a previous result reached for polymeric AM based on the same constant per-layer failure probability $p_{Constant}$ (Baumers and Holweg, 2016). A further major share of cost (27%) arises through labour, which can be split into a small cost impact arising from machine setup, initial supervision and direct post-processing (2%) and a more substantial cost incurred through manual post processing following the build operation (25%). This underlines the point made by Rickenbacher et al. (2013) regarding the significance of considering the full chain of process steps for a realistic perspective. Finally, with a share of around 0.3%, process energy consumption, excluding the energy consumption of all ancillary systems and activities, results in a negligible cost impact.

The next step in the analysis is to use the cost model to explore the effects of build composition on the cost of the AM route. This is done by increasing the number of blower units in each build and estimating $C_{Total}$. With the above described attribute of freely configurable build space, the realistic scenario is to populate the remaining build volume with reference components after the desired number of blower components has been inserted, we refer to such configurations as "synthetic builds". To ensure mixed builds, the build volume packing and time estimation tool was configured to insert at least one unit of each reference part. This approach resulted in 13 iterations of the model, containing 1-13 blowers.

To compare the realistic setting to an analysis of the cost of the blower components manufactured without other components added, the tool was re-executed excluding the reference parts. Mirroring the modelling approach taken in some items of literature (Hopkinson and Dickens, 2003, Ruffo and Hague, 2007; Piili et al., 2015), the estimation tool was able to insert a total of 20 blowers until the build space was exhausted, resulting in 20 iterations of the tool without additional reference parts. We consider the full range of insertible blowers as we are interested in identifying the minimal cost configuration. The results of these executions are presented in Table 4.

**Table 4: Cost modelling results, DMLS pathway**

| Number of units | $v$ | $V_{Build}$ | $T_{Build}$ | Deposition rate | $C_{Build}$ | $C_{Unit}$ | $C_{Total}$ |
|---|---|---|---|---|---|---|---|
| | | | Synthetic builds, containing blowers and representative parts | | | | |
| 1 blower | 1.66% | 505.42 cm$^3$ | 53.38 h | 9.47 cm$^3$/h | € 1,665.63 | € 41.72 | € 56.32 |
| 2 blowers | 1.75% | 479.06 cm$^3$ | 50.99 h | 9.39 cm$^3$/h | € 1,591.72 | € 41.95 | € 56.67 |
| 3 blowers | 1.80% | 465.86 cm$^3$ | 49.82 h | 9.35 cm$^3$/h | € 1,555.23 | € 42.08 | € 56.87 |
| 4 blowers | 1.80% | 466.52 cm$^3$ | 49.87 h | 9.35 cm$^3$/h | € 1,556.84 | € 42.07 | € 56.85 |
| 5 blowers | 2.05% | 409.08 cm$^3$ | 44.71 h | 9.15 cm$^3$/h | € 1,396.98 | € 42.72 | € 57.85 |
| 6 blowers | 2.20% | 382.71 cm$^3$ | 42.34 h | 9.04 cm$^3$/h | € 1,323.50 | € 43.09 | € 58.41 |
| 7 blowers | 2.17% | 386.89 cm$^3$ | 42.71 h | 9.06 cm$^3$/h | € 1,334.94 | € 43.02 | € 58.31 |
| 8 blowers | 2.14% | 393.45 cm$^3$ | 43.29 h | 9.09 cm$^3$/h | € 1,352.93 | € 42.92 | € 58.16 |
| 9 blowers | 1.88% | 447.94 cm$^3$ | 48.16 h | 9.30 cm$^3$/h | € 1,504.18 | € 42.24 | € 57.12 |
| 10 blowers | 1.99% | 421.57 cm$^3$ | 45.79 h | 9.21 cm$^3$/h | € 1,430.67 | € 42.54 | € 57.58 |
| 11 blowers | 2.06% | 408.37 cm$^3$ | 44.60 h | 9.16 cm$^3$/h | € 1,393.77 | € 42.71 | € 57.83 |
| 12 blowers | 2.20% | 382.00 cm$^3$ | 42.23 h | 9.05 cm$^3$/h | € 1,320.27 | € 43.07 | € 58.38 |
| 13 blowers | 2.25% | 373.02 cm$^3$ | 41.41 h | 9.01 cm$^3$/h | € 1,295.08 | € 43.20 | € 58.58 |
| | | | Single geometry, only blower | | | | |
| 1 blower | 100.00% | 8.40 cm$^3$ | 8.86 h | 0.95 cm$^3$/h | € 284.27 | € 298.30 | € 448.19 |
| 2 blowers | 50.00% | 16.80 cm$^3$ | 9.60 h | 1.75 cm$^3$/h | € 307.43 | € 167.74 | € 248.79 |
| 3 blowers | 33.33% | 25.21 cm$^3$ | 10.35 h | 2.44 cm$^3$/h | € 330.58 | € 124.22 | € 182.33 |
| 4 blowers | 25.00% | 33.61 cm$^3$ | 11.09 h | 3.03 cm$^3$/h | € 353.74 | € 102.46 | € 149.09 |
| 5 blowers | 20.00% | 42.01 cm$^3$ | 11.83 h | 3.55 cm$^3$/h | € 376.89 | € 89.41 | € 129.15 |
| 6 blowers | 16.67% | 50.41 cm$^3$ | 12.58 h | 4.01 cm$^3$/h | € 400.05 | € 80.70 | € 115.86 |
| 7 blowers | 14.29% | 58.82 cm$^3$ | 13.32 h | 4.41 cm$^3$/h | € 423.20 | € 74.49 | € 106.36 |
| 8 blowers | 12.50% | 67.22 cm$^3$ | 14.07 h | 4.78 cm$^3$/h | € 446.36 | € 69.82 | € 99.24 |
| 9 blowers | 11.11% | 75.62 cm$^3$ | 14.81 h | 5.11 cm$^3$/h | € 469.51 | € 66.20 | € 93.70 |
| 10 blowers | 10.00% | 84.02 cm$^3$ | 15.56 h | 5.40 cm$^3$/h | € 492.67 | € 63.30 | € 89.27 |
| 11 blowers | 9.09% | 92.42 cm$^3$ | 16.30 h | 5.67 cm$^3$/h | € 515.82 | € 60.92 | € 85.65 |
| 12 blowers | 8.33% | 100.83 cm$^3$ | 17.04 h | 5.92 cm$^3$/h | € 538.98 | € 58.94 | € 82.63 |
| 13 blowers | 7.69% | 109.23 cm$^3$ | 17.79 h | 6.14 cm$^3$/h | € 562.13 | € 57.27 | € 80.07 |
| 14 blowers | 7.14% | 117.63 cm$^3$ | 18.53 h | 6.35 cm$^3$/h | € 585.28 | € 55.84 | € 77.88 |
| 15 blowers | 6.67% | 126.03 cm$^3$ | 19.28 h | 6.54 cm$^3$/h | € 608.44 | € 54.59 | € 75.98 |
| 16 blowers | 6.25% | 134.44 cm$^3$ | 20.02 h | 6.71 cm$^3$/h | € 631.59 | € 53.50 | € 74.32 |
| 17 blowers | 5.88% | 142.84 cm$^3$ | 20.77 h | 6.88 cm$^3$/h | € 654.75 | € 52.54 | € 72.85 |
| 18 blowers | 5.56% | 151.24 cm$^3$ | 21.51 h | 7.03 cm$^3$/h | € 677.90 | € 51.69 | € 71.55 |
| 19 blowers | 5.26% | 159.64 cm$^3$ | 22.25 h | 7.17 cm$^3$/h | € 701.06 | € 50.93 | € 70.38 |
| 20 blowers | 5.00% | 168.04 cm$^3$ | 23.00 h | 7.31 cm$^3$/h | € 724.21 | € 50.24 | € 69.33 |

As can be seen from the upper section of Table 4 for synthetically configured builds, unit cost ($C_{Unit}$ and $C_{Total}$) does not decrease monotonously when the number of blowers contained in the build volume is increased. In fact, the lowest unit cost is observed in the first iteration of the model, with only one blower present ($C_{Total}$= € 56.32). The observed variation (s.d. = 0.73) forms an outcome of the algorithmic selection and insertion of indivisible reference components and should therefore be ignored. Thus, the model suggests that the total unit cost in DMLS is indeed independent of the number of blowers inserted into the build volume, as long as the unused capacity is filled by other parts, as reported by Company C.

The cost model based on synthetically filled builds can be contrasted with the results of the blower-only case (lower section of Table 4). These iterations exhibit a behaviour of decreasing unit cost as the quantity of blowers increases, from initially € 448.19 with a single unit to € 69.33 with 20 blowers contained in the build at maximum capacity. This conforms to the pattern described by Ruffo and Hague (2007) – which is shown to be the outcome of an inability to fill the available build space and therefore lacking realism.

The results reached in this paper are graphically summarised in Figure 6. The inter-process cost comparison between the conventional welding, CNC machining and turning pathway and the DMLS route shows that AM adoption is likely to lead to a unit cost saving of 36% to 46%, depending on process setup. Moreover, Figure 6 shows that the unit cost calculations based on a single geometry overstate the unit cost estimations based on synthetic builds in any case, even if the available capacity is fully utilised (13 units in the synthetic case versus 20 units in the blower-only model). For iterations containing the same number of blowers, the blower-only modelling approach exhibits a mean overstatement of $C_{Total}$ of 157%.

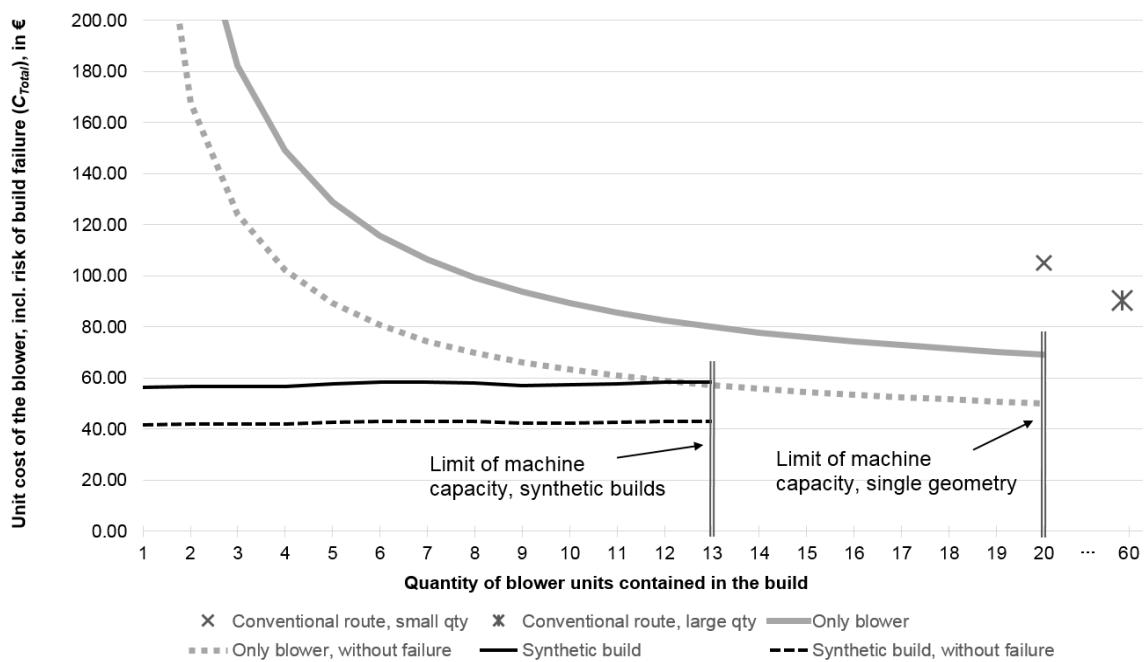

**Figure 6: Inter-process comparison of unit costs**

Providing additional context, the data collected for this research allowed an estimation of the use-phase impact in terms of reduction of energy consumption arising during the blower's useful life. Effectively modelling the energy consumption savings as a continuous annuity, the use-phase savings resulting from reduced energy consumption is estimated using equation (6), assuming an annual discounting rate $r$ of 2% and a useful life $k$ of 7.411 years.

Using these parameters, the calculation of $DS_{Energy}$ indicates a total discounted saving per unit on use-phase energy costs attributable to re-design of € 1980.62. This shows that the estimated downstream saving from the adoption of the AM route is rather large relative to the manufacturing cost saving. Switching from a high volume conventional route (60 units) to the AM route reported by Company C (4 blowers in a mixed build) a manufacturing unit cost reduction of € 34.12 is estimated.

**Discussion**

The empirical results on cost performance reached in this paper indicate that, despite the risks of build failure and the costs of ancillary processes, the AM route appears highly attractive. In terms of the investigated unit costs, this research estimates a unit cost saving of 37.5% when switching from the conventional pathway (low cost) to the AM route as used by Company C. This result should be qualified by stating that this model omits the costs impact of component redesign and re-validation, which may be significant (Mellor et al., 2014).

An additional simplification is that the specification of build failure as a function of the number of deposited layers is reductive and the use of build failure rates taken from polymeric AM systems may be inaccurate. Furthermore, since the blower is the tallest geometry considered in this paper and is present in every investigated build, the probability of build failure is effectively constant. In reality, however, one would expect a relationship between the geometry contained in the build volume and the likelihood of build failure.

This analysis demonstrates that it is possible to model the full unit cost of adopting an AM route despite not knowing the content of each build. For most ex-ante investigations of AM manufacturing cost, where full information on build composition is not available, this approach substantially increases realism. We have shown that the difference between the unit cost estimates originating from synthetically composed builds and those focussing only on one geometry can be substantial.

The results reached by our model can be contextualised in the literature by assessing the specific cost levels resulting from the investigated minimum cost configurations, which is € 6.70 / cm³ for the synthetically populated mixed build and € 8.25 / cm³ for the blower-only case. A head-on comparison with the specific cost outcomes contained in the literature for similar metallic powder bed fusion systems is presented in Table 5. Our estimated cost level for the blower only case is identical to that estimated by Piili et al. (2015), at € 8.25 / cm³, who omit risk of build failure and pre-and post-processing however. The configuration of our model corresponding to the study by Rickenbacher et al. (2013) is the blower-only case with 5 units contained in the build, resulting in a specific cost of € 15.37 / cm³. We note that Rickenbacher et al. report a very high machine cost rate of € 90.00 / h which is more than five times the machine cost rate measured in our research (€ 17.66 / h), thereby explaining the very high specific cost level reported.

**Table 5: Specific cost results reported in the literature on metallic AM**

| Reference | System and material grade | Notable cost model elements | Specific cost estimate |
|---|---|---|---|
| Atzeni and Salmi (2012) | EOSINT M270, AlSi10Mg | <ul><li>Multi-part build, single geometry</li><li>Includes pre-and post processing (heat treatment)</li></ul> | € 7.92 /cm³ * |
| Baumers et al., (2013) | EOSINT M270, Stainless steel 316L | <ul><li>Multi-part build, multiple geometries</li><li>Including wire erosion to separate parts from build plate</li></ul> | € 7.03 /cm³ † |
| Rickenbacher et al. (2013) | Unspecified selective laser melting system, unspecified material grade | <ul><li>Multi-part build, single geometry</li><li>Includes pre-and post-processing, including wire erosion to separate substrate</li></ul> | € 28.20 /cm³ to € 66.75 /cm³ * |
| Piili et al. (2015) | Unspecified selective laser melting prototype system, Stainless steel PH1 | <ul><li>Multi-part build, single geometry</li></ul> | € 8.25 /cm³ * |

\* = specific cost not cited explicitly; inferred from the data provided

† = currency converted using historical exchange rate from 1 January of reference year

In terms of broader implications, our results show that, at sub-maximal levels of capacity utilisation, unit costs are dependent on quantity as well as build composition, as expressed by Ruffo and Hague (2007). Yet, if it is considered that the units of build space are capable of mutual substitution in AM, known as the property of "fungibility", AM users are in principle able to drive up the degree of capacity utilisation through populating available machine capacity by inserting other parts, subject to manufacturability constraints. This observation suggests that to make any statement on efficient AM utilisation, as required by technology adoption decisions, a problem of using the available capacity must be addressed. Whether this problem extends to other aspects such as machine scheduling forms a subject for future investigation.

The redesigned blower was investigated primarily to ensure technical feasibility in a study of unit cost of different manufacturing processes. In the frame of our cost model we expect that design parameters have an effect on the cost performance of the system if changes in the degree of capacity utilisation alter indirect cost levels, changes to material and energy inputs affect direct costs, and if design changes result in larger or smaller post-processing requirements. Further, we note that the scope of our analysis has excluded the consideration of lead time, design costs, inspection, and in-process inventory, which all form topics for future studies.

The inclusion of a simple analysis of use-phase benefits underlines that follow-on considerations may outweigh manufacturing considerations in manufacturing: after all, the adoption of a new manufacturing technology should manifest itself in better products (Stoneman, 2002). Centring on use-phase fractional energy consumption associated with the blower module, the projected use-phase cost saving far outweighs the estimated manufacturing cost saving; the use of AM thus is associated with important value creation for the final user. As stated by the manufacturer of the packaging machine (Company A), a conscious decision was made to not charge an increased price for the system, despite such improved product performance. Assuming that the blowers are sourced by Company A at cost, the share of value capture $\theta$ on the unit level resulting from AM technology adoption between Company A ($\theta$) and the end user (1- $\theta$) can thus be estimated:

$$\theta = \frac{C_{Conventional,High} - C_{Total}}{DS_{Energy} + (C_{Conventional,High} - C_{Total})} = 1.69\% \qquad (7)$$

This result suggests that Company A currently passes on the value increase resulting from AM technology adoption almost entirely to the end customer. It is important to note, however, that additional supply chain costs in terms of warehousing and logistics are not considered in this paper.

**Conclusions**

In this paper we have extended existing activity-based costing methodologies to inform an inter-process comparison between a conventional pathway, combining a subtractive process and welding, and an AM route utilising the AM technology variant DMLS. The approach carries novelty by combining diverse aspects into a new, more realistic, model of AM costs. The main aspects are: (1) representation of AM as a chain of processes, (2) determination of efficient build configurations including reference parts to fill excess capacity, (3) the expected cost effect of the possibility of build failure and (4) design adaptation to ensure technical validity in an inter-process comparison. The build time estimator used in this model has been validated by Baumers et al. (2013) and we have compared the resulting specific cost levels to those reported in the existing literature, indicating robustness of our model.

In proposing this methodology we depart from existing AM cost models, where it is simply assumed that the available build space is efficiently used (Achillas et al, 2017;

Atzeni et al., 2010; Atzeni and Salmi, 2012; Hopkinson and Dickens, 2003; Piili et al., 2015; Rickenbacher et al., 2013) or where the unused capacity is left empty (Alexander et al., 1998; Ruffo and Hague, 2007). Following the notion that in AM empty capacity can be allocated to other jobs or sold to outside bidders, this paper moves away from treating the build capacity as indivisible and constrained to one particular production run.

By introducing the characteristic of fungibility and showing that realistic cost estimates can be constructed on the basis of synthetically specified build configurations, the developed model suggests that the relationship between quantity and total unit cost is immaterial in AM if build space can be filled otherwise. Of course, throughput-borne economies of scale continue to exist in AM (Baumers et al., 2016) as only "static" economies of scale associated with the indivisibility of tooling are absent (cf. Haldi and Whitcomb, 1967).

However, the characteristic of fungibility does not only have operational consequences. As discussed by Gilder (1990) in the context of photolithographic processes used in the manufacture if integrated circuitry, the free-of-cost realisation of additional ("marginal") logical structures indicates the lack of a relationship between the performance of a design and its unit cost. Baumers et al. (2016) have extended this argument to AM by correlating process energy consumption and product shape complexity. Within the limitations of our costing framework and concentrating on the core AM process and a limited number of pre- and post-processing steps, we argue that specifics of geometry, and hence product functionality, are not related to unit cost if capacity utilisation, raw material usage and pre- and post-processing are unaffected. This assumption forms the basis for claims that AM adoption will make considerable additional design space accessible (Rosen et al., 2007; De Mul, 2016).

By incorporating a simple model of use-phase savings arising from AM technology adoption, this paper has also provided a glimpse into the relationship between cost savings originating in different stages of the product lifecycle. By specifying a value share parameter, it has been shown for the investigated application that manufacturing cost savings can be far outweighed by use-phase cost saving resulting from efficiency gains. Our case study therefore demonstrates that use-phase considerations are important and should not be ignored in technology adoption decisions surrounding AM.

**Acknowledgements**

The authors wish to express their gratitude to Stefano Boccolari and Claudio Ferrari, Manufacturing Technology Manager and Manufacturing Technology Specialist at Tetra Pak for their support in providing the data flowing into this model. Additionally, the authors wish to thank Simone Casella, General Manager at Beam-IT and Alessandro Vecchini, Technical Manager at CALF.

# References

- Achillas, C., Tzetzis, D. and Raimondo, M.O., 2017. Alternative production strategies based on the comparison of additive and traditional manufacturing technologies. International Journal of Production Research, pp.1-13.

- Alexander, P., Allen, S. and Dutta, D., 1998. Part orientation and build cost determination in layered manufacturing. Computer-Aided Design, 30(5), pp.343-356.

- Araújo, L.J.P., Özcan, E., Atkin, J.A., Baumers, M., Tuck, C. and Hague, R., 2015. Toward Better Build Volume Packing In Additive Manufacturing: Classification Of Existing Problems And Benchmarks.

- Atzeni, E., Iuliano, L., Minetola, P. and Salmi, A., 2010. Redesign and cost estimation of rapid manufactured plastic parts. Rapid Prototyping Journal, 16(5), pp.308-317.

- Atzeni, E. and Salmi, A., 2012. Economics of additive manufacturing for end-usable metal parts. The International Journal of Advanced Manufacturing Technology, 62(9-12), pp.1147-1155.

- Baumers, M., Dickens, P., Tuck, C. and Hague, R., 2016. The cost of additive manufacturing: machine productivity, economies of scale and technology-push. Technological Forecasting and Social Change, 102, pp.193-201.

- Baumers, M., and Holweg, M.,2016. Cost Impact of the Risk of Build Failure in Laser Sintering. Proceedings of the Solid Freeform Fabrication Symposium 2016, University of Texas at Austin.

- Baumers, M., Tuck, C., Wildman, R., Ashcroft, I., Rosamond, E. and Hague, R., 2013. Transparency Built-in. Journal of Industrial Ecology, 17(3), pp.418-431.

- Baumers, M., Tuck, C., Wildman, R., Ashcroft, I. and Hague, R., 2011. Energy inputs to additive manufacturing: does capacity utilization matter? Proceedings of the Solid Freeform Fabrication Symposium 2011, University of Texas at Austin

- Beltrametti, L., Gasparre, A., 2016. The Adoption of Additive Technologies in Manufacturing: an Economic and Organizational Outlook. Cavallari, M., Harfouche A. (edited by) The Social Relevance of the Organisation of Information Systems and ICT, Springer, Berlin (forthcoming).

- Berman, B., 2012. 3-D printing: The new industrial revolution. Business horizons, 55(2), pp.155-162.

- Berumen, S., Bechmann, F., Lindner, S., Kruth, J.P. and Craeghs, T., 2010. Quality control of laser-and powder bed-based Additive Manufacturing (AM) technologies. Physics procedia, 5, pp.617-622.

- Boothroyd, G., Dewhurst, P., and Knight, W., 1994. Product Design for Manufacture and Assembly. New York; Marcel Dekker.

- EOS GmbH, 2016. Corporate website. Available at: http://www.eos.info/en [Accessed 04/09/2016]

- Eyers, D.R., Wong, H., Wang, Y. and Dotchev, K., 2008, September. Rapid manufactured enabled mass customisation: untapped research opportunities in supply chain management. In A. Lyons ed.,, Proceedings of the Logistics Research Network Annual Conference, Liverpool, UK (pp. 10-12).

- Gardan, J., 2016. Additive manufacturing technologies: State of the art and trends. International Journal of Production Research, 54(10), pp.3118-3132.

- Gilder, G., 1990. Microcosm: the quantum revolution in economics and technology. Simon and Schuster.

- Haldi, J. and Whitcomb, D., 1967. Economies of scale in industrial plants. The Journal of Political Economy, pp.373-385.

- Holmström, J., Liotta, G. and Chaudhuri, A., 2017. Sustainability outcomes through direct digital manufacturing-based operational practices: A design theory approach. Journal of Cleaner Production.

- Hopkinson, N. and Dicknes, P., 2003. Analysis of rapid manufacturing—using layer manufacturing processes for production. Proceedings of the Institution of Mechanical Engineers, Part C: Journal of Mechanical Engineering Science, 217(1), pp.31-39.

- Khorram Niaki, M. and Nonino, F., 2017. Additive manufacturing management: a review and future research agenda. International Journal of Production Research, 55(5), pp.1419-1439.

- Li, Y., Jia, G., Cheng, Y. and Hu, Y., 2017. Additive manufacturing technology in spare parts supply chain: a comparative study. International Journal of Production Research, 55(5), pp.1498-1515.

- Materialise NV, 2017. Corporate website. Available at: http:// http://www.materialise.com/en/home [Accessed 01/03/2017]

- Mellor, S., Hao, L. and Zhang, D., 2014. Additive manufacturing: A framework for implementation. International Journal of Production Economics, 149, pp.194-201.

- de Mul, J., 2016. Possible Printings: On 3D Printing, Database Ontology, and Open (Meta) Design. In 3D Printing (pp. 87-98). TMC Asser Press.

- Niazi, A., Dai, J.S., Balabani, S. and Seneviratne, L., 2006. Product cost estimation: Technique classification and methodology review. Journal of manufacturing science and engineering, 128(2), pp.563-575.

- Norman, G. and Thisse, J.F., 1999. Technology choice and market structure: strategic aspects of flexible manufacturing. The Journal of Industrial Economics, 47(3), pp.345-372.

- Piili, H., Happonen, A., Väistö, T., Venkataramanan, V., Partanen, J. and Salminen, A., 2015. Cost Estimation of Laser Additive Manufacturing of Stainless Steel. Physics Procedia, 78, pp.388-396.

- Reuter, M., 2007. Methodik der Werkstoffauswahl. Hanser, Munich.

- Rickenbacher, L., Spierings, A. and Wegener, K., 2013. An integrated cost-model for selective laser melting (SLM). Rapid Prototyping Journal, 19(3), pp.208-214.

- Rosen, D.W., 2007, August. Design for additive manufacturing: a method to explore unexplored regions of the design space. In Eighteenth Annual Solid Freeform Fabrication Symposium (pp. 402-415).

- Ruffo, M. and Hague, R., 2007. Cost estimation for rapid manufacturing - simultaneous production of mixed components using laser sintering. Proceedings of the Institution of Mechanical Engineers, Part B: Journal of Engineering Manufacture, 221(11), pp.1585-1591.

- Schmid, M. and Levy, G., 2014. Quality management and estimation of quality costs for additive manufacturing with SLS. ETH-Zürich.

- Son, Y.K., 1991. A cost estimation model for advanced manufacturing systems. The International Journal of Production Research, 29(3), pp.441-452.

- Stoneman, P., 2001. The economics of technological diffusion. Wiley-Blackwell.

- Tuck, C., Hague, R. and Burns, N., 2006. Rapid manufacturing: impact on supply chain methodologies and practice. International journal of services and operations management, 3(1), pp.1-22.

- Weller, C., Kleer, R. and Piller, F.T., 2015. Economic implications of 3D printing: market structure models in light of additive manufacturing revisited. International Journal of Production Economics, 164, pp.43-56.